\documentclass[twoside,prl,twocolumn,amsfonts,amssymb,amsmath,lengthcheck,showpacs]{revtex4}
\usepackage{textcomp}
\usepackage{graphicx}
\usepackage{color}

\newcommand{\bra}[1]{\ensuremath{\left\langle #1 \right|}}
\newcommand{\ket}[1]{\ensuremath{\left| #1\right\rangle}}

\renewcommand{\mod}[1]{\ensuremath{\left|#1\right|}}

\renewcommand{\subsection}[1]{\vspace{5pt}{\centering\bf #1\\}\vspace{5pt}}

\begin{document}

\title{Final-state effects in the radio frequency spectrum of strongly interacting fermions}
\date{\today}
\author{Sourish Basu}
\email{sourish@ccmr.cornell.edu}
\author{Erich J. Mueller}
\email{emueller@ccmr.cornell.edu}
\affiliation{Laboratory of Atomic and Solid State Physics, Cornell University, Ithaca, New York 14853}
\begin{abstract}
We model the impact of final state interactions on the radio frequency spectrum of  a strongly interacting two-component superfluid Fermi gas.  In addition to a broad asymmetric peak coming from the break-up of Cooper pairs we find that, for appropriate parameters, one can observe a sharp symmetric ``bound-bound" spectral line coming from  the conversion of Cooper pairs in one channels to pairs/molecules in another.
\end{abstract}
\maketitle
Radio frequency (RF) spectroscopy may become a powerful probe of the many-body state of a gas of cold atoms.  One indicator of this potential is  the sharpness of hyperfine spectral lines \cite{narrow}: orders of magnitude smaller than the  $\sim$100 kHz interaction strengths found in interacting clouds of lithium atoms.  As cold atom experiments begin probing exotic states of matter, this separation of energy scales may allow one to detect subtle atomic correlations.

Despite this optimistic viewpoint, there appears to be major theoretical holes in our understanding of RF spectra, even at a qualitative level.  The primary difficulty is that when radio waves change the hyperfine spin of an atom, the entire many-body state needs to adjust.  Including these final state interactions is a nontrivial many-body problem, akin to the one which must be solved to understand the X-ray spectra of metals \cite{combescot}.
These are hard problems -- there is no generic prescription for including these final state interactions, rather the solution will depend on the system at hand. Here we present a variational calculation in which we calculate the radio frequency spectrum of a  strongly interacting superfluid two-component Fermi gas, including arbitrarily strong short range interactions in the final state. We find structure in the spectra which have not previously been theoretically described.

We consider a model where a gas of neutral Fermions occupy two different hyperfine states  \ket{1} and \ket{2}. Radio waves drive a transition from \ket{2} to a new hyperfine state \ket{3}. Interactions between atoms in the three states are described by scattering lengths $a_{12}$, $a_{13}$, and $a_{23}$.

Previous approaches to understanding the RF spectrum of a superfluid Fermi gas have either neglected final state interactions \cite{Torma-2004, Ohashi-Griffin-2005, Levin-2005,Levin-2007, massignan}, or included them with sum rules \cite{Baym-2006,Baym-2007,Zwerger-2007}, diagramatics \cite{perali-RF}, or energy arguments \cite{muellertrap}.  Many of these previous works focussed on trying to gain quantitative understanding of the spectra for very specific parameter values. Here we present a straightforward variational approach which complements these other works: we focus on gaining a global picture of the qualitative structure of the RF spectrum for all parameter values.  While elementary, our formalism is quite powerful -- as was kindly pointed out to us by Giancarlo Strinati, it is equivalent to the zero-temperature limit of the BCS-RPA theory used by Perali et al. \cite{perali-RF} to explore the unitary limit.

Final state interactions can qualitatively change the RF spectrum. For example, early experiments \cite{insensitive} with the three lowest hyperfine states of Lithium atoms \cite{LithiumStates} found that final state interactions nearly canceled out the contribution from interactions in the initial state. This is due to the proximity of three wide Feshbach resonances: $B_{12}=834$G ($\Delta B_{12} = 300$G), $B_{13}=690$G ($\Delta B_{13} = 122$G) and $B_{23}=811$G ($\Delta B_{23} = 222$G). The scattering lengths, which diverge on resonance, all become large at the same time. When $a_{12}$ and $a_{13}$ are similar the interaction effects cancel \cite{Baym-2006}. Later experiments  \cite{Jin-2005,chingap,Chin-Julienne-2005,ketterlepairs,Shin:05-25-07} found  nontrivial spectra -- some of which were used as evidence of pairing in these gases.

Here we use a BCS ansatz for the initial state, and a restricted set of final states which includes only ``coherent" excitations: the quasihole created in the $|2\rangle$ state is restricted to have the same momentum as the excited atom. The neglected excitations will shift our spectral lines and introduce broadening, however our approach should capture the qualitative features of the spectrum.  Our approach is exact in the limiting case of $a_{12}=a_{13}$, where the spectrum is a single delta-function peak, unshifted from its free space value. Furthermore, when $a_{13}=0$, it reduces to the BCS result: a broad asymmetric peak corresponding to the breaking up of Cooper pairs. Generically we find a bimodal spectrum containing both the delta-function and the broad continuum, though for some parameter ranges the delta-function lies in the continuum and becomes significantly broadened -- disappearing as $a_{13}\to0$. These results mirror those found by Chin and Julienne in studying the RF spectrum of a single molecule \cite{Chin-Julienne-2005}. Interpreting the delta function peak as a ``bound-bound" transition,  we encounter final state pairs in parameter ranges where no free-space molecules exist.

\begin{figure}[tbp]
 \includegraphics[width=\columnwidth]{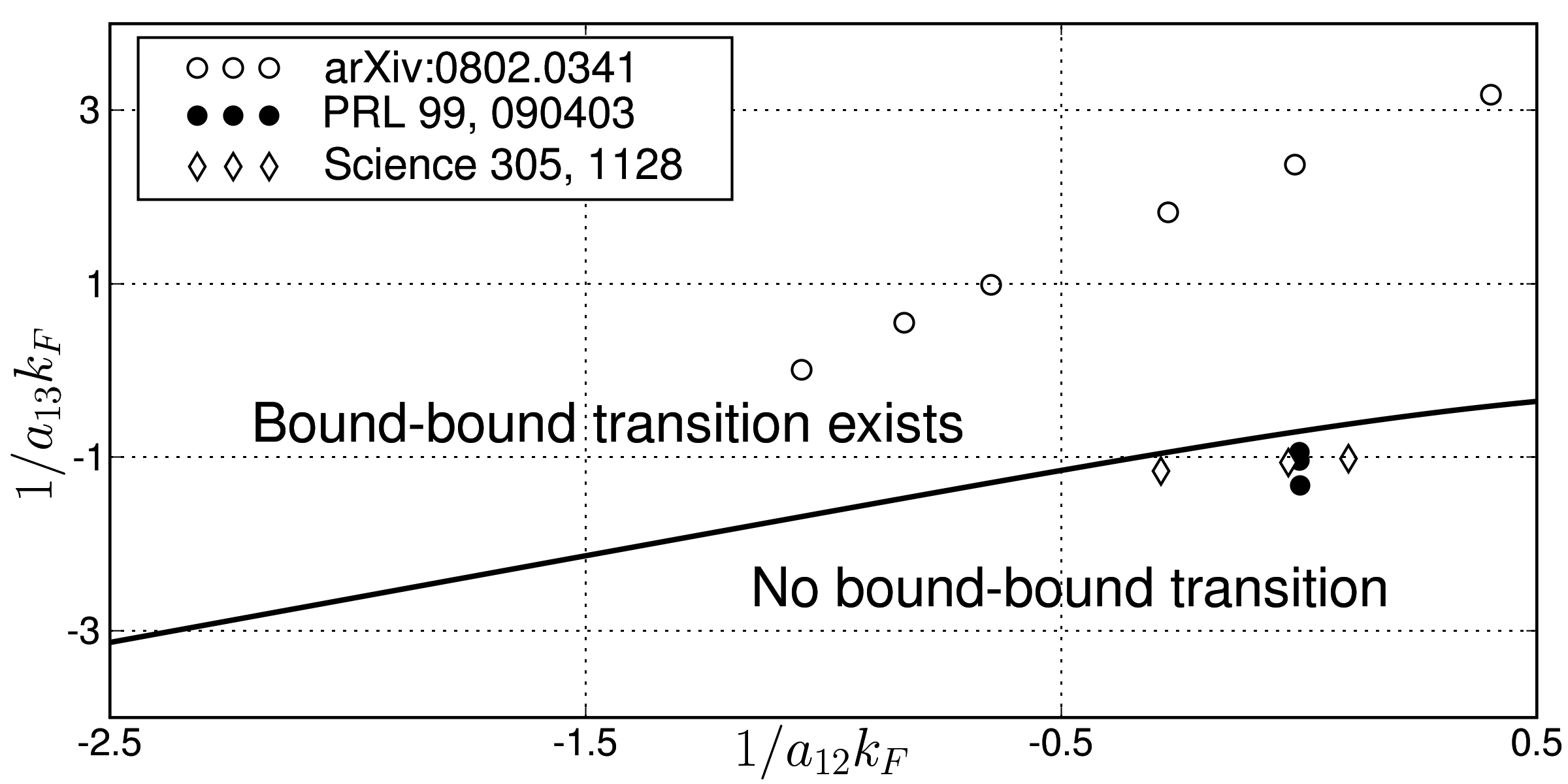}
 \caption{\label{fig:fig1} 
Phase diagram showing for which values of initial ($a_{12}$) and final ($a_{13}$) scattering lengths the homogeneous RF spectrum contains a bound-bound peak.  Scattering lengths are normalized via the fermi wavevector, $k_F = (3\pi^2 n)^{1/3}$, where $n$ is the atomic density.  Symbols: parameters from lowest temperature data in the experiments -- references \cite{chingap}, \cite{Shin:05-25-07} and \cite{Ketterle:10-07} correspond to open diamonds, closed circles, and open circles respectively.}
\end{figure}

We construct a ``phase diagram'' (figure~\ref{fig:fig1}) in the final interaction ($a_{13}k_F$) versus initial interaction ($a_{12}k_F$) plane, delineating the regions where we expect to see a bound-bound transition and where we do not. 
We also plot the interaction values used in three experiments \cite{Shin:05-25-07,Ketterle:10-07,chingap}.  The experimental observations are consistent with our predictions; in \cite{Shin:05-25-07} no bound-bound transition is seen, while in \cite{Ketterle:10-07} one is always seen.  

We only consider the spectrum of the homogeneous gas (measurable tomographically \cite{Shin:05-25-07}), and avoid discussing the trap averaged spectrum, where inhomogeneous broadening obscures the basic physics \cite{muellertrap}. Furthermore, we restrict ourselves to the unpolarized case ($n_1=n_2$) at $T=0$.

The role of the hyperfine energies $\epsilon_j$ (j=1,2,3), is most transparent if we make the Canonical transformation to the field operators $\psi_{j,k}\to e^{-i\epsilon_j t}\psi_{j,k}$, to arrive at a Hamiltonian
${\cal H} = H + H_{\rm RF}$ with $H=H_1+H_2+H_3+H_{12}+H_{13}$,
\begin{eqnarray}
 H_j &=& \sum_{k} \epsilon_k\psi^\dag_{j,k}\psi_{j,k} \\
 H_{ij}&=&-\lambda_{ij} \sum_{kpq} \psi^\dag_{i,k}\psi^\dag_{j,p}\psi_{j,p-q}\psi_{i,k+q}\\
  H_{\rm RF} &=& \sum_k \left( \psi^\dag_{3,k}\psi_{2,k} e^{-i\omega t} + \psi^\dag_{2,k}\psi_{3,k} e^{i\omega t} \right)\\\nonumber
& =& e^{-i\omega t} X + e^{i\omega t} X^\dag,
\end{eqnarray}
We neglect the $H_{23}$ term\cite{H23neglect}. The RF detuning is $\omega=\omega_{RF}-(\epsilon_3-\epsilon_2)$ and the free particle dispersion is  $\epsilon_k=k^2/2m-\mu$, where $m$ is the atomic mass, and we work in units where $\hbar=1$.  The interaction parameters $\lambda_{ij}$ are connected to the scattering lengths $a_{ij}$ between states \ket{i} and \ket{j} by 
\begin{equation}
 \frac{1}{\lambda_{ij}} = -\frac{mV}{4\pi a_{ij}} + \sum_k \frac{m}{k^2}
\end{equation}
where $V$ is the spatial volume. 
Of particular importance is that if $\lambda_{12}=\lambda_{13}$, then $[X,H]=0$, implying that the RF spectrum will consist of a single unshifted delta-function.

\begin{figure*}[htbp]
 \includegraphics[width=\textwidth]{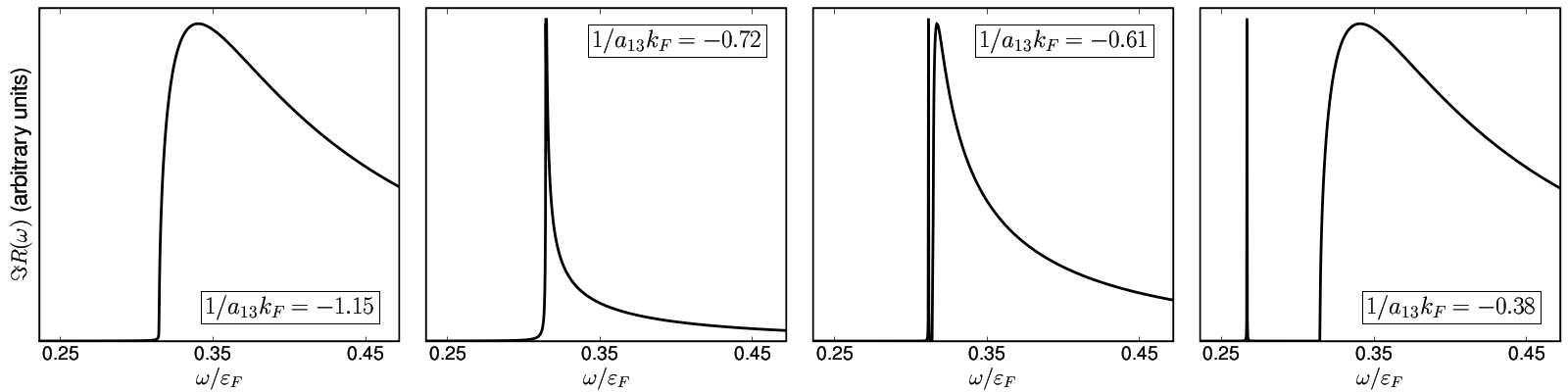}
 \caption{\label{fig:fig2} Theoretical radio frequency spectra at $a_{12}\rightarrow\infty$, where $k_F = (3\pi^2 n)^{1/3}$ and $\epsilon_F=k_f^2/2m$. The discrete peak on the right two figures correspond to a "bound-bound" transition between Cooper pairs in the various channels.  Moving towards the left, that peak merges with the continuum. The height of the discrete peak has been scaled down for display; the actual spectral weight in the peak is shown in figure~\ref{fig:fig4}.}
\end{figure*}

We make the variational ansatz that the system is initially in the BCS state 
\begin{equation}
\ket{\rm GS}=\prod_k \left(u_k+ v_k \psi_{1,k}^\dagger \psi_{2,-k}^\dagger\right) \ket{\rm vac}.
\end{equation}
Minimizing $\langle H-\mu N\rangle$, and neglecting the Hartree terms \cite{Diener-2008} yields a ground state energy $E_{GS}=\langle H\rangle=\sum_k (\epsilon_k-E_k) + \Delta^2/\lambda_{12}$, with $v_k^2=(E_k-\epsilon_k)/2E_k$, and $u_k^2=(E_k+\epsilon_k)/2E_k$.  The gap $\Delta$ obeys the equation
\begin{equation}
\lambda_{12}\sum_k\frac{1}{2E_k} = 1
\end{equation}
and the number of particles is $N=2\sum_k |v_k|^2$.
The excitations will be described by quasiparticle operators $\gamma_k=u_k \psi_{1,k} - v_k \psi_{2,-k}^\dag$ and $\eta_k = v_k \psi^\dag_{1,k} + u_k\psi_{2,-k}$ with energies $E_k=\sqrt{(\epsilon_k-\mu)^2+\Delta^2}$.

Fermi's Golden rule states that  given an RF detuning $\omega$, the rate of transitions $I$ to a set of final states \ket{f} is given by the imaginary part of a spin response function,
\begin{eqnarray}\label{eq:resp}
I(\omega)&\propto&\sum_f \mod{\bra{GS}H_{\rm RF}\ket{f}}^2 \delta(\omega+ E_{\rm GS} - E_f)\\\nonumber
&=&{\rm Im} \bra{\rm GS} X^\dag \frac{1}{\omega-\bar H} X \ket{\rm GS}={\rm Im} R(\omega),
\end{eqnarray}
where $\bar H=H-E_{GS}$.
We approximate this response by restricting the final states to those containing a single quasiparticle.  By definition this captures the ``coherent" part of the response where most of the spectral weight lies \cite{Pethick:013618}, and aside from the BCS ansatz is exact when $a_{13}=a_{12}$ or when $a_{13}=0$.
Introducing the {\em incomplete} but orthornormal set of states
\begin{equation}
 \ket{k} = \frac{1}{v_k}\psi_{3,k}^\dag \psi_{1,k}\ket{\rm GS} = -\psi^\dag_{3,k}\gamma^\dag_{-k}\ket{\rm GS},
\end{equation}
we replace the $\bar H$ in \eqref{eq:resp} by $\tilde H=\sum_{k,k^\prime}\ket{k}\bar H_{kk^\prime}\bra{k^\prime}$, where  $\bar H_{kk^\prime}=\bra{k}\bar H\ket{k^\prime}  = \delta_{kk^\prime}\left(E_k+\epsilon_k\right) - \lambda_{13} u_k u_{k^\prime}$.  Matrix inversion gives 
\begin{eqnarray}\label{eq:rullr}
 \frac{R(\omega)}{V}&=& \frac{1}{V}
 \sum_{k,k^\prime} v_k v_{k^\prime} \left[(\omega-\tilde H)^{-1}\right]_{kk^\prime}\\\nonumber
 &=& \frac{(2m\mu)^{3/2}}{(2\pi)^2\mu}\left[G_3 + \frac{2}{\pi}\frac{\bar{\Delta}^2 G_2^2}{\frac{1}{a_{13}\sqrt{2m\mu}}+\frac{2}{\pi}G_1}\right].
 \end{eqnarray}
 In terms of scaled variables
 $\bar{\omega}=\omega/\mu$, $\bar{\Delta}=\Delta/\mu$ and $E_x=\sqrt{(x^2-1)^2+\bar{\Delta}^2}$,
 \begin{eqnarray}\nonumber
 G_1 &=& \int_0^\infty dx \left[-1-\frac{x^2(E_x+x^2-1)}{E_x(\bar{\omega}-E_x-x^2+1)}\right]\\
  G_2 &=& \int_0^\infty \frac{x^2 dx}{E_x(\bar{\omega}-x^2-E_x+1)}\\\nonumber
  G_3 &=& \int_0^\infty x^2 dx\frac{E_x-x^2+1}{E_x(\bar{\omega}-E_x-x^2+1)}.
\end{eqnarray}

As previously discussed, there are two contributions to the spectrum: a continuum from breaking up Cooper pairs, and a discrete peak from the conversion of a Cooper pair into a pair in the new channel. Mathematically, the continuum comes from the poles of the integrands of the $G$'s.  Since we have severely restricted the available final states and neglected Hartree terms, the location of this continuum is independent of the final-state interactions and corresponds to all $\omega$ for which we can find a $k$ with $\omega=E_k+\epsilon_k$: {\em ie.} {$\omega>\sqrt{\Delta^2+\mu^2}-\mu$}. In a more sophisticated theory the threshold will depend on $a_{13}$.
The discrete peak comes from the condition
\begin{equation}\label{dgap}
\frac{1}{a_{13}\sqrt{2m\mu}}+\frac{2}{\pi}G_1=0.
\end{equation}
This condition can only be satisfied for sufficiently strong interactions.  In particular, if $a_{12}=\infty$, we find that this peak exists if  $a_{13}>0$ or if $a_{13}\sqrt{2m \mu}\lesssim -1$.  In the former case the excited state can be thought of as a molecular pair, but in the latter case it is more analogous to a Cooper pair.  When $a_{12}=a_{13}$, one can recognize that if one sets $\omega=0$ in Eq.~(\ref{dgap}) 
one recovers the gap equation.  In this limit all spectral weight resides in this zero-detuning peak. 

\begin{figure}[htbp]
 \includegraphics[width=\columnwidth]{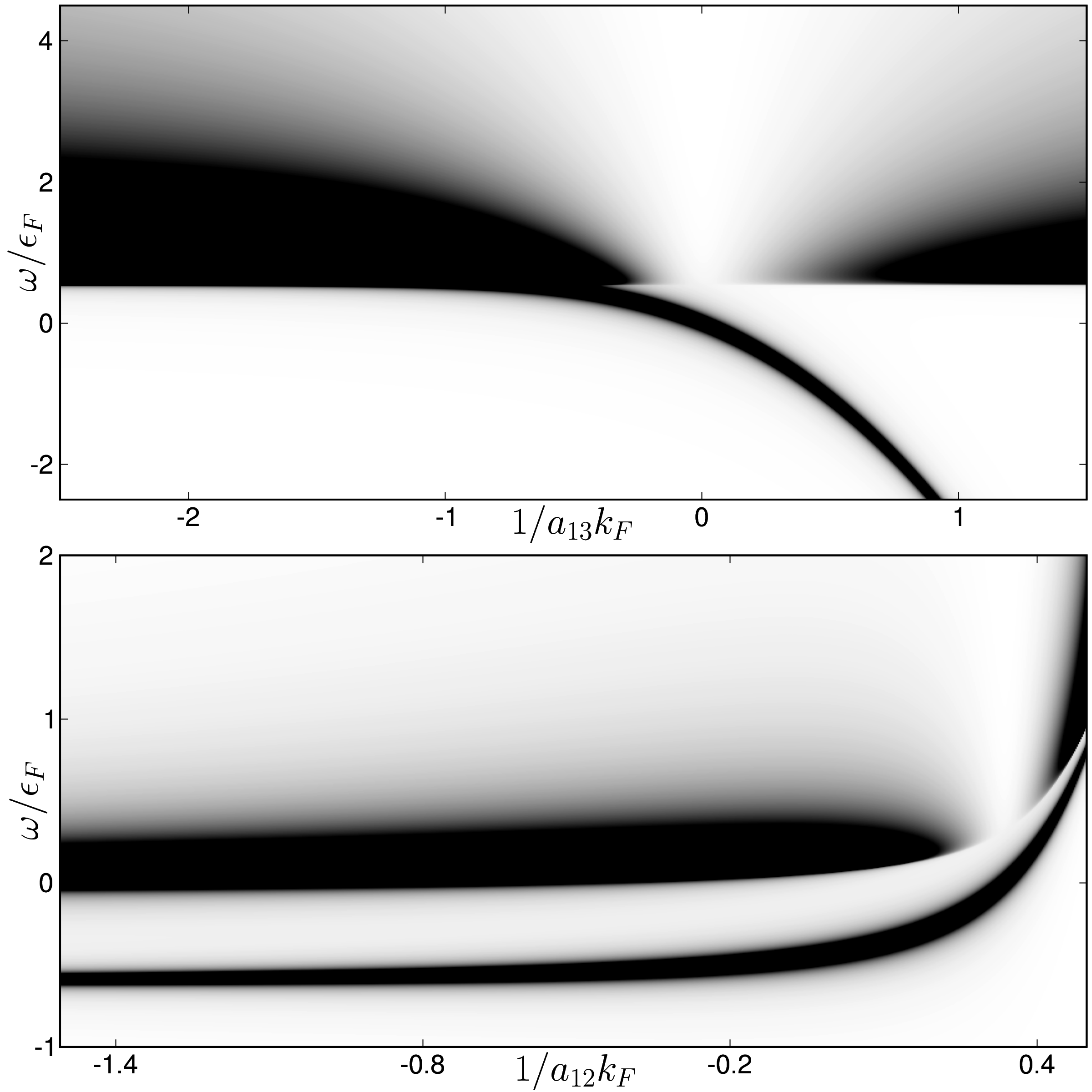}
 \caption{\label{fig:fig3}\textbf{[Top]} Spectral density as a function of final state scattering length $a_{13}$ and RF frequency $\omega$, when initial state interactions satisfy $a_{12}=\infty$.
 % and Fermi energy $\epsilon_F=k_F^2/2m$
   \textbf{[Bottom]} Same plot, except as a function of initial state scattering length $a_{12}$ and $\omega$ with $a_{13}=\infty$. Spectral weight increases from white to black.}
\end{figure}

\begin{figure}[htbp]
 \includegraphics[width=\columnwidth]{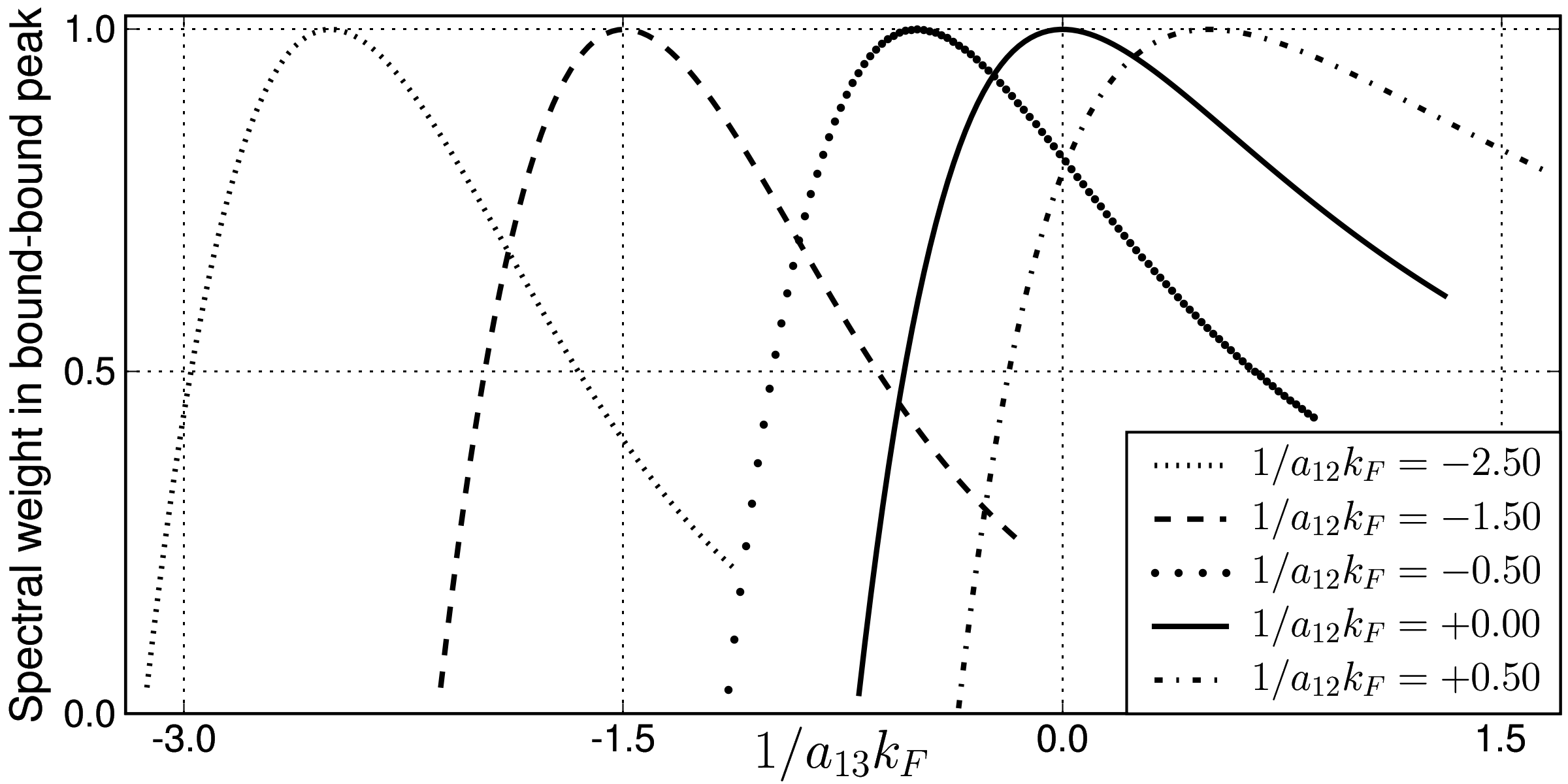}
 \caption{\label{fig:fig4} Fraction of spectral weight in the bound-bound peak as a function of final state interaction strength $a_{13}$, for different values of $a_{12}$. When the two interactions are equal, i.e., $a_{13}=a_{12}$, all the weight is in this peak.}
\end{figure}

In the BCS-BEC crossover, the chemical potential depends on interactions and density in a nontrivial way.  Following convention we will quote our results in terms of the Fermi momentum of an ideal Fermi gas with the same density as the one we are considering, $k_F = (3\pi^2 n)^{1/3}$. In figures~\ref{fig:fig2} and \ref{fig:fig3} (top) we illustrate the $a_{13}$ dependence of the spectrum by showing the spectral density in the case where $a_{12}=\infty$.
As revealed by figures~\ref{fig:fig1}, \ref{fig:fig2} and \ref{fig:fig3}, the pair-pair transition peak breaks off from the continuum around $a_{13}k_F=-0.7$.  As depicted by the solid line in figure~\ref{fig:fig4}, most of the spectral weight is in that peak close to unitarity. For $0<a_{13}k_F\ll 1$ (beyond the right hand edge of figures \ref{fig:fig3} (top) and \ref{fig:fig4}) the spectral weight shifts back to the continuum. In that regime, the wavefunction overlap between the large \ket{1}-\ket{2} Cooper pairs and the small \ket{1}-\ket{3} molecules becomes negligible. Further, the spectral weight in the bound-bound peak goes to zero when it hits the continuum (left edge of the curves in figure~\ref{fig:fig4}), in agreement with \cite{Chin-Julienne-2005}. Figure~\ref{fig:fig3} (bottom) is similar to figure~\ref{fig:fig3} (top), except that we show the $a_{12}$ dependence of the spectrum keeping $a_{13}=\infty$.

We now discuss how the structure seen in figures \ref{fig:fig2} and \ref{fig:fig3} can be measured experimentally.  First, since all physics only depends on the dimensionless parameters $a_{ij} k_F$ all interactions can be modified by changing the density.  If $a_{12}=\infty$ then {\em only} the final state interaction is changed when the density is varied.  This is a quite powerful knob:  as seen in figure~\ref{fig:fig1}, two of the experiments \cite{Shin:05-25-07,chingap} have been performed very close to the point in Fig.~\ref{fig:fig3} (top) where a bound-bound transition emerges from the continuum.  A modest increase in density should allow the observation of this feature.  Alternatively, one can produce different final state interactions by changing which internal states one uses in the experiment \cite{Ketterle:10-07}.  Restricting ourselves to Lithium \cite{LithiumStates} in states \ket{1}, \ket{2} and \ket{3}, one has six combinations:  for example one could take a \ket{1}\ket{2} superfluid and excite \ket{2} to \ket{3}, or one could take a \ket{2}\ket{3} superfluid and excite \ket{2} to \ket{1}.  Using the notation that $ij-k$ represents a \ket{i}\ket{j} superfluid where one excites \ket{j} to \ket{k}, tuning the $ij$ interaction to resonance, and taking typical central densities \cite{Shin:05-25-07}, typical final state interactions ($1/a_{ik}k_F$) are \cite{scattheory}
12-3:-0.94,
21-3:-0.19,
31-2:2.58,
13-2:2.14,
32-1:-0.86,
23-1:0.18.

Given the focus on sum rules in recent theoretical studies \cite{Baym-2006,Baym-2007,Zwerger-2007}, we compute the zeroth and first moments of the resolvent  $R_{\rm exact}$, defined in eq~\eqref{eq:resp}, and our approximation,
 $R_{\rm approx}$, from
  eq~\eqref{eq:rullr} . The $|k\rangle$ basis is sufficiently large that the zeroth moment, or total spectral weight, is preserved by our approximation,
\begin{equation}
 \begin{split}
  \int_{-\infty}^\infty \frac{d\omega}{\pi} \Im R_{\rm exact}(\omega) &= \bra{\rm GS}X^\dag X\ket{\rm GS}
  = \sum_k v_k^2\\
  \int_{-\infty}^\infty \frac{d\omega}{\pi} \Im R_{\rm approx}(\omega) &= \sum_k \mod{\bra{k}X\ket{\rm GS}}^2
  = \sum_k v_k^2
 \end{split}
\end{equation}
The higher moments of both $R_{\rm exact}$ and $R_{\rm approx}$, on the other hand, are dominated by long tails which arise from the short distance structure of the BCS pairs. In the absence of final state interaction, for example, $R_{\rm approx}(\omega)\propto \omega^{-3/2}$ for large $\omega$. With final state interaction, $R_{\rm approx}(\omega)\propto \omega^{-5/2}$ and our approach reproduces the first spectral moment calculated in \cite{Baym-2006}.

Finally we comment on the quantitative accuracy of our approach.  The largest approximation that we make is the neglect of Hartree-type interactions.  These terms will shift the spectral lines.  We also neglected incoherent processes which will broaden the spectra.  Another important approximation was our neglect of finite temperature effects.  These could be important in experiments where $T/T_F\sim 0.1$.  Despite the severity of these approximations, we believe that our calculation is valuable for its simplicity. It illustrates the important physics in a transparent manner.

We would like to thank Tin-Lun Ho, Mohit Randeria, and Wilhelm Zwerger for motivational comments, Wolfgang Ketterle for discussions about the experiments, Kaden Hazzard for various insights, and Kathy Levin for correcting our sum rule arguments. We found discussions with Henk Stoof valuable for thinking about diagrammatic approaches to this problem.  We are particularly indebted to Giancoli Strinati who sent us detailed notes showing how to connect this variational approach to the BCS-RPA.
This material is based upon work supported by the National Science Foundation through grants PHY-0456261 and PHY-0758104.


\begin{thebibliography}{99}
\bibitem{LithiumStates}
 \ket{1}, \ket{2}, \ket{3}=  \ket{1/2,1/2}, \ket{1/2,-1/2}  \ket{3/2,-3/2} where  \ket{F,m_F} are their zero-field quantum numbers
\bibitem{H23neglect}
By the Pauli exclusion principle, the \ket{2} atoms are spatially anticorrelated.  To lowest order, the RF excitation will create \ket{3} atoms which share these correlations, rendering negligible the impact of  $H_{23}$.
\end{thebibliography}
\end{document}